\documentclass[12pt]{article}
\newcommand{\be}{\begin{equation}}
\newcommand{\ee}{\end{equation}}
\newcommand{\ba}{\begin{eqnarray}}
\newcommand{\ea}{\end{eqnarray}}
\newcommand{\no}{\nonumber\\}
\begin{document}

\title{General formulae for $f_1 \to f_2 \gamma$}
\author{L.\ Lavoura \\
\small Universidade T\'ecnica de Lisboa \\
\small Centro de F\'\i sica das Interac\c c\~oes Fundamentais \\
\small Instituto Superior T\'ecnico, 1049-001 Lisboa, Portugal}
\date{25 February 2003}
\maketitle

\begin{abstract}
At one-loop level the decay $f_1 \to f_2 \gamma$,
where $f_1$ and $f_2$ are two spin-$1/2$ particles
with the same electric charge,
is mediated by a boson $B$ and a spin-$1/2$ fermion $F$.
The boson $B$ may have either spin $0$ --- interacting with the fermions
through Dirac matrices $1$ and $\gamma_5$ --- or spin $1$ --- with
$V+A$ and $V-A$ couplings to the fermions.
I give general formulae
for the one-loop electroweak amplitude of $f_1 \to f_2 \gamma$
in all these cases.
\end{abstract}

\section{Introduction}

Radiative decays like $\mu \to e \gamma$
and $b \to s \gamma$ provide an important testing ground
for many models in particle physics.
In particular,
the experimental bounds on flavour-changing leptonic radiative decays
\cite{bounds} are planned to improve,
in some cases by a few orders of magnitude \cite{better},
and the relevance of those decays in tests of new physics
will certainly increase.

It is important for model builders
to be able to compute expeditiously the predictions of their models
for radiative decays.
Unfortunately,
QCD effects are important and blur the picture
in hadronic decays like $b \to s \gamma$.
On the other hand,
in flavour-changing leptonic decays only the electroweak theory is relevant,
and simple,
closed formulae may be produced.

The amplitude for $\mu \to e \gamma$ in the standard electroweak theory
with lepton mixing (either light or heavy neutrinos)
has been given by Cheng and Li \cite{book}.
However,
that amplitude has been computed
for gauge bosons with exclusively left-handed interactions.
Recently,
the same authors together with He \cite{chengli}
have computed the amplitude for $\mu \to e \gamma$
following from a general Yukawa interaction.

In this paper I give simple formulae for the amplitude of
$f_1 \to f_2 \gamma$ following from
either a general (axial-)vector interaction
or a general Yukawa interaction.
My formulae are more general than the ones given in the references above,
since
\begin{itemize}
\item I allow for arbitrary electric charges of the fermions $f_1$ and $f_2$,
and of the internal fields --- a fermion $F$ and a boson $B$ --- in
the one-loop diagram responsible for the decay;
\item I do not neglect the masses of $f_1$ and $f_2$
in the loop integrals;
\item I allow for a general gauge interaction,
with both $V-A$ and $V+A$ components.
\end{itemize}
The last point is important
since gauge bosons displaying $V+A$ interactions are present in many theories.
In particular,
\begin{itemize}
\item In the left--right-symmetric model \cite{lrsm}
there is a charged gauge boson
$W_R^\pm$ coupling to the fermions like $V+A$ and,
as a matter of fact,
the observed $W^\pm$ is supposed to have a small $W_R^\pm$ component;
\item In models
with vector-like fermions \cite{vectorlike} --- like for instance
the $E_6$ grand unified theory,
which has both vector-like charge-$-1/3$ quarks
and vector-like charge-$-1$ leptons --- the neutral gauge bosons
couple to flavour-changing currents
while retaining both $V-A$ and $V+A$ couplings;
\item In the 3-3-1 model \cite{331},
based on the electroweak gauge group $SU(3) \times U(1)$,
both singly and doubly charged vector bosons exist,
and they have both $V-A$ and $V+A$ couplings to the fermions.
\end{itemize}

The one-loop computation of $f_1 \to f_2 \gamma$ is non-trivial
since there are both vertex-type diagrams --- in which
the photon attaches to either the internal boson $B$
or the internal fermion $F$ --- and self-energy-type diagrams --- in which
the photon attaches to either $f_1$ or $f_2$.
One must write the (divergent) two-point integrals
in terms of three-point integrals
in order to be able to add the diagrams of both types.
When one does that one finds that the full vertex is both gauge-invariant
and finite,
as it ought to be.

The plan of this paper is as follows.
In section 2 I give the notation for the gauge-invariant amplitude.
In section 3 I define the relevant three-point finite loop integrals
in terms of which the amplitude will be written.
In section 4 I give the amplitude resulting from the Yukawa couplings
to a spin-$0$ boson.
In section 5 I give the amplitude following from
the couplings of the fermions to an intermediate vector boson.

\section{Notation for the vertex}

I want to compute the process
$f_1 \left( p_1 \right) \to f_2 \left( p_2 \right) \gamma \left( q \right)$,
where $q = p_1 - p_2$.
The fermion $f_1$ has mass $m_1$ while $f_2$ has mass $m_2$.
The fermions are on mass shell:
$p_1^2 = m_1^2$ and $p_2^2 = m_2^2$.
The fermions $f_1$ and $f_2$
are represented by spinors $u_1$ and $\bar u_2$,
respectively,
which satisfy $p_1 \hskip-10pt / \,\, u_1 = m_1 u_1$
and $\bar u_2\, p_2 \hskip-10pt / = m_2 \bar u_2$.

The amplitude for the decay is $e \epsilon_\mu^\ast \left( q \right) M^\mu$,
where $\epsilon_\mu^\ast \left( q \right)$
is the polarization vector of the outgoing photon
and $e$ is the electric charge of the positron.
Gauge invariance implies that $q_\mu M^\mu$ must be zero;
therefore $M^\mu$ must be of the form
\be
M^\mu = \bar u_2 \left( \sigma_L \Sigma_L^\mu + \sigma_R \Sigma_R^\mu
+ \delta_L \Delta_L^\mu + \delta_R \Delta_R^\mu \right) u_1\, ,
\ee
where $\sigma_L$,
$\sigma_R$,
$\delta_L$,
and $\delta_R$ are numerical coefficients with dimension of inverse mass,
and
\ba
\Sigma_L^\mu &=& \left( p_1^\mu + p_2^\mu \right) \gamma_L
- \gamma^\mu \left( m_2 \gamma_L + m_1 \gamma_R \right),
\label{sigmaL} \\
\Sigma_R^\mu &=& \left( p_1^\mu + p_2^\mu \right) \gamma_R
- \gamma^\mu \left( m_2 \gamma_R + m_1 \gamma_L \right),
\label{sigmapR} \\
\Delta_L^\mu &=& q^\mu \gamma_L + \frac{q^2}{m_2^2 - m_1^2}\, \gamma^\mu
\left( m_2 \gamma_L + m_1 \gamma_R \right),
\\
\Delta_R^\mu &=& q^\mu \gamma_R + \frac{q^2}{m_2^2 - m_1^2}\, \gamma^\mu
\left( m_2 \gamma_R + m_1 \gamma_L \right).
\ea
The matrices
$\gamma_L = \left. \left( 1 - \gamma_5 \right) \right/ 2$
and $\gamma_R = \left. \left( 1 + \gamma_5 \right) \right/ 2$
are the projectors of chirality.
If we define $\sigma^{\mu \nu} = \left( i / 2 \right)
\left[ \gamma^\mu, \gamma^\nu \right]$,
then $M^\mu$ may alternatively be written as
\be
M^\mu = \bar u_2 \left[ i\sigma^{\mu \nu} q_\nu
\left( \sigma_L \gamma_L + \sigma_R \gamma_R \right)
+ \delta_L \Delta_L^\mu + \delta_R \Delta_R^\mu \right] u_1\, .
\ee

Only the coefficients $\sigma_L$ and $\sigma_R$ are relevant
to the physical decay $f_1 \to f_2 \gamma$,
because $\epsilon_\mu^\ast \left( q \right) q^\mu = 0$
and $q^2 = 0$ for an on-shell photon.
The coefficients $\delta_L$ and $\delta_R$ are important
when $f_1 \left( p_1 \right)
\to f_2 \left( p_2 \right) \gamma \left( q \right)$
is just a sub-process of a more complex decay,
like for instance $f_1 \left( p_1 \right) \to f_2 \left( p_2 \right) e^+ e^-$.
In this paper I shall only give $\sigma_L$ and $\sigma_R$.
The partial width for $f_1 \to f_2 \gamma$ is
\be
\Gamma = \frac{\left( m_1^2 - m_2^2 \right)^3
\left( \left| \sigma_L \right|^2 + \left| \sigma_R \right|^2 \right)}
{16 \pi m_1^3}\, .
\ee

\section{The basic integrals} \label{passarino}

The expressions for the coefficients $\sigma_L$ and $\sigma_R$
will be given in terms of a few loop integrals.
Denote
\ba
D_B &=& k^2 - m_B^2\, , \label{hfurt} \\
D_{1F} &=& \left( k + p_1 \right)^2 - m_F^2\, , \\
D_{2F} &=& \left( k + p_2 \right)^2 - m_F^2\, .
\ea
Then,
I define
\ba
a &=& \int \! \frac{d^4 k}{\left( 2 \pi \right)^4}\,
\frac{1}{D_B D_{1F} D_{2F}}\, , \hspace*{3mm} \\
c_1 p_1^\theta + c_2 p_2^\theta &=&
\int \! \frac{d^4 k}{\left( 2 \pi \right)^4}\,
\frac{k^\theta}{D_B D_{1F} D_{2F}}\, , \\
d_1 p_1^\theta p_1^\psi +
d_2 p_2^\theta p_2^\psi +
f \left( p_1^\theta p_2^\psi + p_2^\theta p_1^\psi \right)
+ x g^{\theta \psi}
&=& \int \! \frac{d^4 k}{\left( 2 \pi \right)^4}\,
\frac{k^\theta k^\psi}{D_B D_{1F} D_{2F}}\, .
\ea
In the formulae for $\sigma_{L,R}$ only the finite coefficients $a$,
$c_1$,
$c_2$,
$d_1$,
$d_2$,
and $f$ occur;
the divergent $x$ cancels out with two-point integrals.

Conversely,
let
\ba
D_{1B} &=& \left( k - p_1 \right)^2 - m_B^2\, , \\
D_{2B} &=& \left( k - p_2 \right)^2 - m_B^2\, , \\
D_F &=& k^2 - m_F^2\, .
\ea
Then,
\ba
\bar a &=& \int \! \frac{d^4 k}{\left( 2 \pi \right)^4}\,
\frac{1}{D_{1B} D_{2B} D_F}\, , \hspace*{3mm} \\
\bar c_1 p_1^\theta + \bar c_2 p_2^\theta &=&
\int \! \frac{d^4 k}{\left( 2 \pi \right)^4}\,
\frac{k^\theta}{D_{1B} D_{2B} D_F}\, , \\
\bar d_1 p_1^\theta p_1^\psi + \bar d_2 p_2^\theta p_2^\psi
+ \bar f \left( p_1^\theta p_2^\psi+ p_2^\theta p_1^\psi \right)
+ \bar x g^{\theta \psi} &=&
\int \! \frac{d^4 k}{\left( 2 \pi \right)^4}\,
\frac{k^\theta k^\psi}{D_{1B} D_{2B} D_F}\, . \label{nfgty}
\ea

When one uses the approximation $m_1^2 = m_2^2 = 0$
together with $q^2 = 0$ the integrals may be computed easily.
Defining $t = m_F^2 / m_B^2$,
one obtains
\ba
a &=& \frac{i}{16 \pi^2 m_B^2} \left[
\frac{- 1}{t - 1}
+ \frac{\ln{t}}{\left( t - 1 \right)^2} \right], \label{a} \\
c_1 = c_2 \equiv c &=& \frac{i}{16 \pi^2 m_B^2} \left[
\frac{t - 3}{4 \left( t - 1 \right)^2}
+ \frac{\ln{t}}{2 \left( t - 1 \right)^3} \right], \\
d_1 = d_2 = 2 f \equiv d &=& \frac{i}{16 \pi^2 m_B^2} \left[
\frac{- 2 t^2 + 7 t - 11}{18 \left( t - 1 \right)^3}
+ \frac{\ln{t}}{3 \left( t - 1 \right)^4} \right], \\
\bar a &=& \frac{i}{16 \pi^2 m_B^2} \left[
\frac{1}{t - 1}
- \frac{t \ln{t}}{\left( t - 1 \right)^2} \right], \\
\bar c_1 = \bar c_2 \equiv \bar c &=& \frac{i}{16 \pi^2 m_B^2} \left[
\frac{3 t - 1}{4 \left( t - 1 \right)^2}
- \frac{t^2 \ln{t}}{2 \left( t - 1 \right)^3} \right], \\
\bar d_1 = \bar d_2 = 2 \bar f \equiv \bar d
&=& \frac{i}{16 \pi^2 m_B^2} \left[
\frac{11 t^2 - 7 t + 2}{18 \left( t - 1 \right)^3}
- \frac{t^3 \ln{t}}{3 \left( t - 1 \right)^4} \right]. \label{bard}
\ea

\section{Results for a Yukawa interaction}

The fermions $f_1$ and $f_2$
may have an Yukawa interaction with a spin-$0$ boson $B$
and with another spin-$1/2$ fermion $F$,
assumed to be distinct from both $f_1$ and $f_2$.
Let us write that interaction as
\be
\mathcal{L}_{\rm Yukawa} = \sum_{i=1}^2 \left[
B \bar F \left( L_i \gamma_L + R_i \gamma_R \right) f_i
+ B^\ast \bar f_i \left( L_i^\ast \gamma_R
+ R_i^\ast \gamma_L \right) F \right],
\label{uifgz}
\ee
with arbitrary dimensionless numerical coefficients $L_1$,
$L_2$,
$R_1$,
and $R_2$.
I denote
\ba
\lambda &=& L_2^\ast L_1\, , \label{lambda} \\
\rho &=& R_2^\ast R_1\, , \\
\zeta &=& L_2^\ast R_1\, , \\
\upsilon &=& R_2^\ast L_1\, \label{upsilon} .
\ea

The electric charges of $f_1$ and $f_2$,
in units of $e$,
are $Q_f$;
the electric charge of $F$ is $Q_F$
and the electric charge of $B$ is $Q_B$.
Obviously,
from eq.~(\ref{uifgz}),
\be
Q_f = Q_F - Q_B\, . \label{kwlcm}
\ee
Otherwise I allow for arbitrary $Q_f$,
$Q_F$,
and $Q_B$.

Let us consider the consequences of the Yukawa interaction in eq.~(\ref{uifgz})
for the vertex $f_1 \left( p_1 \right)
\to f_2 \left( p_2 \right) \gamma \left( q \right)$.
There will in general be four diagrams for that vertex:
two self-energy diagrams in which the photon attaches
either to $f_1$ or to $f_2$;
one diagram in which the photon attaches to $F$;
and another diagram in which the photon attaches to $B$.
The self-energy diagrams are proportional to $Q_f$,
and the other two diagrams are proportional to $Q_F$ and $Q_B$,
respectively.
One uses eq.~(\ref{kwlcm})
to write the vertex as the sum of two terms,
one of them proportional to $Q_F$
and the other one proportional to $Q_B$.

The mass of the scalar boson $B$ is denoted $m_B$
and the mass of the fermion $F$ is denoted $m_F$.
With the loop integrals defined in the previous section I construct
\ba
k_1 &=& m_1 \left( c_1 + d_1 + f \right), \label{k1} \\
k_2 &=& m_2 \left( c_2 + d_2 + f \right), \\
k_3 &=& m_F \left( c_1 + c_2 \right),
\ea
and
\ba
\bar k_1 &=& m_1 \left( - \bar c_1 + \bar d_1 + \bar f \right), \\
\bar k_2 &=& m_2 \left( - \bar c_2 + \bar d_2 + \bar f \right), \\
\bar k_3 &=& m_F \left( - \bar a + \bar c_1 + \bar c_2 \right).
\label{bark3}
\ea
The results for $\sigma_L$ and $\sigma_R$
are written in terms of these functions:
\ba
\sigma_L &=&
Q_F \left( \rho k_1 + \lambda k_2 + \upsilon k_3 \right)
+ Q_B \left( \rho \bar k_1 + \lambda \bar k_2 + \upsilon \bar k_3 \right),
\label{jjjj1} \\
\sigma_R &=&
Q_F \left( \lambda k_1 + \rho k_2 + \zeta k_3 \right)
+ Q_B \left( \lambda \bar k_1 + \rho \bar k_2 + \zeta \bar k_3 \right).
\label{sigmaR}
\ea

The results in eqs.~(\ref{k1})--(\ref{sigmaR})
do not involve any approximations
and they are fully general --- they hold even when the photon is off-shell,
$q^2 \neq 0$.
One may want to keep the mass prefactors in the $k_1, k_2, \ldots,
\bar k_3$ of eqs.~(\ref{k1})--(\ref{bark3}),
while computing $c_1 + d_1 + f, c_2 + d_2 + f, \ldots,
- \bar a + \bar c_1 + \bar c_2$
in the approximation $m_1^2 = m_2^2 = 0$ (and $q^2 = 0$).
One uses eqs.~(\ref{a})--(\ref{bard}) and obtains
\ba
\left( - i \right) 16 \pi^2 m_B^2 \left( c + \frac{3}{2}\, d \right)
&=& \frac{t^2 - 5 t - 2}{12 \left( t - 1 \right)^3}
+ \frac{t \ln{t}}{2 \left( t - 1 \right)^4}\, ,
\label{hr} \\
\left( - i \right) 16 \pi^2 m_B^2
\left( - \bar c + \frac{3}{2}\, \bar d \right)
&=& \frac{2 t^2 + 5 t - 1}{12 \left( t - 1 \right)^3}
- \frac{t^2 \ln{t}}{2 \left( t - 1 \right)^4}\, ,
\label{gr} \\
\left( - i \right) 16 \pi^2 m_B^2 \left( 2 c \right) &=&
\frac{t - 3}{2 \left( t - 1 \right)^2}
+ \frac{\ln{t}}{\left( t - 1 \right)^3}\, ,
\label{kr} \\
\left( - i \right) 16 \pi^2 m_B^2 \left( - \bar a + 2 \bar c \right) &=&
\frac{t + 1}{2 \left( t - 1 \right)^2}
- \frac{t \ln{t}}{\left( t - 1 \right)^3}\, .
\label{ir}
\ea
The functions in the right-hand sides of eqs.~(\ref{hr})--(\ref{ir})
have been given in ref.~\cite{chengli},
where they were called $H(r)$,
$G(r)$,
$K(r)$,
and $I(r)$,
respectively (with $r = t-1$ and apart from a common factor $2$).
They are all positive definite,
decreasing functions,
which start at $t = 0$ with a value smaller than $1$
and tend to $0$ as $t^{-1}$ when $t \to \infty$.
The exception is the function in the right-hand side of eq.~(\ref{kr}),
which tends to infinity as $- 3/2 - \ln{t}$ in the limit $t \to 0$.

\section{Results for a gauge interaction}

Now suppose that the fermions $f_1$ and $f_2$
interact with a (neutral or charged) vector boson $B_\alpha$
and with another fermion $F$,\footnote{I use the same notation $F$
as in the previous section
for the fermion with which $f_1$ and $f_2$ interact,
although the fermion $F$ will not in general be the same
in the Yukawa interaction and in the gauge interaction.
In the same vein,
I use identical notations $m_{B,F}$ and $Q_{B,F}$
for the masses and electric charges,
respectively,
of the intermediate boson and fermion.}
assumed to be distinct from both $f_1$ and $f_2$,
the interaction Lagrangian being
\be
\mathcal{L}_{\rm gauge} = \sum_{i=1}^2 \left[
B_\alpha \bar F \gamma^\alpha
\left( L_i^\prime \gamma_L + R_i^\prime \gamma_R \right) f_i
+ B_\alpha^\ast \bar f_i \gamma^\alpha
\left( {L_i^\prime}^\ast \gamma_L + {R_i^\prime}^\ast \gamma_R \right)
F \right],
\label{jsoeh}
\ee
with arbitrary dimensionless numerical coefficients $L_1^\prime$,
$L_2^\prime$,
$R_1^\prime$,
and $R_2^\prime$.
I use the notation
\ba
\lambda^\prime &=& {L_2^\prime}^\ast \! L_1^\prime\, , \\
\rho^\prime &=& {R_2^\prime}^\ast \! R_1^\prime\, , \\
\zeta^\prime &=& {L_2^\prime}^\ast \! R_1^\prime\, , \\
\upsilon^\prime &=& {R_2^\prime}^\ast \! L_1^\prime\, .
\ea
The electric charge of $F$ is $Q_F$,
in units of $e$,
and the electric charge of $B_\alpha$ is $Q_B$.
Again,
eq.~(\ref{kwlcm}) holds.
The mass of $B_\alpha$ is $m_B$ and the mass of $F$ is $m_F$.

The massive gauge field $B_\alpha$ has associated with it a scalar
``would-be Goldstone boson'' $\varphi$,
while $\varphi^\ast$ is associated with $B_\alpha^\ast$.
The Yukawa interaction of $f_1$ and $f_2$ with $F$
and with the ``would-be Goldstone bosons'' $\varphi$ and $\varphi^\ast$
is given by\footnote{The phases of $\varphi$ and $\varphi^\ast$
are implicitly defined through eq.~(\ref{idurm}) in a convenient way.}
\ba
\mathcal{L}_\varphi &=&
\varphi\, \frac{i}{m_B}\, \sum_{i=1}^2 \bar F
\left[ \left( R_i^\prime m_i - L_i^\prime m_F \right) \gamma_L
+ \left( L_i^\prime m_i - R_i^\prime m_F \right) \gamma_R \right] f_i
\no & &
+ \varphi^\ast\, \frac{i}{m_B}\, \sum_{i=1}^2 \bar f_i
\left[ \left( {L_i^\prime}^\ast m_F - {R_i^\prime}^\ast m_i \right)
\gamma_R
+ \left( {R_i^\prime}^\ast m_F - {L_i^\prime}^\ast m_i \right)
\gamma_L \right] F\, . \hspace*{6mm}
\label{idurm}
\ea

I assume that,
just as in the Standard Model (SM),
the three-gauge-boson vertex of a photon $A_\mu$ with outgoing momentum $q$,
an incoming $B_\alpha$ with incoming momentum $p$,
and an incoming $B_\beta^\ast$ with incoming momentum $\bar p$
(obviously $p + \bar p = q$)
has the following Feynman rule:
\be
i e Q_B \left[ g^{\alpha \beta} \left( p - \bar p \right)^\mu
- g^{\mu \alpha} \left( q + p \right)^\beta
+ g^{\mu \beta} \left( q + \bar p \right)^\alpha \right].
\ee
Furthermore,
I assume that the vertex of $A_\mu$ with (incoming)
$B_\alpha^\ast$ and $\varphi$
has Feynman rule $e Q_B m_B g^{\mu \alpha}$,
while the vertex of $A_\mu$ with $B_\alpha$ and $\varphi^\ast$
is $- e Q_B m_B g^{\mu \alpha}$.
This is,
once again,
analogous to what happens in the SM.

One adds the contributions from diagrams with $B_\alpha$
with those from diagrams with $\varphi$
and with those from diagrams with both $B_\alpha$ and $\varphi$.
All diagrams must be computed in the same gauge --- I have used
the Feynman--'t Hooft gauge,
in which the propagators of both $B_\alpha$ and $\varphi$
have poles exclusively at the physical mass $m_B$.
One obtains
\ba
\sigma_L &=&
Q_F \left( \rho^\prime y_1 + \lambda^\prime y_2
+ \upsilon^\prime y_3 + \zeta^\prime y_4 \right)
+ Q_B \left( \rho^\prime \bar y_1 + \lambda^\prime \bar y_2
+ \upsilon^\prime \bar y_3 + \zeta^\prime \bar y_4 \right), \hspace*{6mm}
\label{uirmd} \\
\sigma_R &=&
Q_F \left( \lambda^\prime y_1 + \rho^\prime y_2
+ \zeta^\prime y_3 + \upsilon^\prime y_4 \right)
+ Q_B \left( \lambda^\prime \bar y_1 + \rho^\prime \bar y_2
+ \zeta^\prime \bar y_3 + \upsilon^\prime \bar y_4 \right),
\ea
with
\ba
y_1 &=& m_1 \left[ 2 a + 4 c_1  + 2 c_2 + 2 d_1 + 2 f
+ \frac{m_F^2}{m_B^2} \left( - c_2 + d_1 + f \right)
\right. \no & & \left.
+ \frac{m_2^2}{m_B^2} \left( c_2 + d_2 + f \right) \right], \\
y_2 &=& m_2 \left[ 2 a + 2 c_1 + 4 c_2 + 2 d_2 + 2 f
+ \frac{m_F^2}{m_B^2} \left( - c_1 + d_2 + f \right)
\right. \no & & \left.
+ \frac{m_1^2}{m_B^2} \left( c_1 + d_1 + f \right) \right], \\
y_3 &=& m_F \left[ - 4 a - 4 c_1 - 4 c_2
+ \frac{m_F^2}{m_B^2} \left( c_1 + c_2 \right)
\right. \no & & \left.
- \frac{m_1^2}{m_B^2} \left( c_1 + d_1 + f \right)
- \frac{m_2^2}{m_B^2} \left( c_2 + d_2 + f \right) \right], \\
y_4 &=& - \frac{m_1 m_2 m_F}{m_B^2} \left( d_1 + d_2 + 2 f \right),
\ea
and
\ba
\bar y_1 &=& m_1 \left[ 2 \bar c_2 + 2 \bar d_1 + 2 \bar f
+ \frac{m_F^2}{m_B^2} \left( \bar a - 2 \bar c_1 - \bar c_2
+ \bar d_1 + \bar f \right)
\right. \no & & \left.
+ \frac{m_2^2}{m_B^2} \left( - \bar c_2 + \bar d_2 + \bar f \right)
\right], \\
\bar y_2 &=& m_2 \left[ 2 \bar c_1 + 2 \bar d_2 + 2 \bar f
+ \frac{m_F^2}{m_B^2} \left( \bar a - \bar c_1 - 2 \bar c_2
+ \bar d_2 + \bar f \right)
\right. \no & & \left.
+ \frac{m_1^2}{m_B^2} \left( - \bar c_1 + \bar d_1 + \bar f \right)
\right], \\
\bar y_3 &=& m_F \left[ - 4 \bar c_1 - 4 \bar c_2
+ \frac{m_F^2}{m_B^2} \left( - \bar a + \bar c_1 + \bar c_2 \right)
\right. \no & & \left.
+ \frac{m_1^2}{m_B^2} \left( \bar c_1 - \bar d_1 - \bar f \right)
+ \frac{m_2^2}{m_B^2} \left( \bar c_2 - \bar d_2 - \bar f \right) \right], \\
\bar y_4 &=& \frac{m_1 m_2 m_F}{m_B^2}
\left( - \bar a + 2 \bar c_1 + 2 \bar c_2
- \bar d_1 - \bar d_2 - 2 \bar f \right). \label{nfhit}
\ea

Just as in the previous section,
the results in eqs.~(\ref{uirmd})--(\ref{nfhit})
are completely general --- they hold even when $q^2 \neq 0$.
One may want to keep the mass prefactors in $y_1$--$y_3$
and in $\bar y_1$--$\bar y_3$
while computing the functions inside the square brackets
in the limit $m_1^2 = m_2^2 = 0$ (and $q^2 = 0$).
With $t = m_F^2 / m_B^2$ as before,
one obtains
\ba
\left( - i \right) 16 \pi^2 m_B^2
\left[ 2 a + 6 c + 3 d + t \left( - c + \frac{3}{2}\, d \right) \right]
&=& \frac{- 5 t^3 + 9 t^2 - 30 t + 8}{12 \left( t - 1 \right)^3}
\no & & +\frac{3 t^2 \ln{t}}{2 \left( t - 1 \right)^4}\, ,
\label{p1} \\
\left( - i \right) 16 \pi^2 m_B^2
\left[ 2 \bar c + 3 \bar d + t \left( \bar a - 3 \bar c
+ \frac{3}{2}\, \bar d \right) \right]
&=& \frac{- 4 t^3 + 45 t^2 - 33 t + 10}{12 \left( t - 1 \right)^3}
\no & & - \frac{3 t^3 \ln{t}}{2 \left( t - 1 \right)^4}\, ,
\label{p2} \\
\left( - i \right) 16 \pi^2 m_B^2
\left( - 4 a - 8 c + 2 t c \right)
&=& \frac{t^2 + t + 4}{2 \left( t - 1 \right)^2}
\no & & - \frac{3 t \ln{t}}{\left( t - 1 \right)^3}\, ,
\label{p3} \\
\left( - i \right) 16 \pi^2 m_B^2
\left[ - 8 \bar c + t \left( - \bar a + 2 \bar c \right) \right]
&=& \frac{t^2 - 11 t + 4}{2 \left( t - 1 \right)^2}
\no & & + \frac{3 t^2 \ln{t}}{\left( t - 1 \right)^3}\, .
\label{p4}
\ea
The function in the right-hand side of eq.~(\ref{p2})
has been given in ref.~\cite{book};
the functions in the right-hand sides of eqs.~(\ref{p1}),
(\ref{p3}),
and (\ref{p4}) are new.
The functions in eqs.~(\ref{p3}) and (\ref{p4}) are positive definite
and decrease continuously from $2$ at $t=0$ to $1/2$ at $t \to \infty$;
the functions in eqs.~(\ref{p1}) and (\ref{p2}) are negative definite
and increase from a value larger than $-1$ at $t=0$
to a value smaller than $0$ at $t \to \infty$.

\section{Conclusions}

The amplitude for the decay $f_1 \to f_2 \gamma$
involves two relevant operators,
$\Sigma_L^\mu$ and $\Sigma^\mu_R$
given in eqs.~(\ref{sigmaL}) and (\ref{sigmapR}).
When $f_1$ and $f_2$ interact with a scalar boson $B$ and with
a fermion $F$ ($F \neq f_1$ and $F \neq f_2$) as in eq.~(\ref{uifgz}),
the coefficients of those operators in the amplitude,
$\sigma_L$ and $\sigma_R$,
receive contributions as in eqs.~(\ref{k1})--(\ref{sigmaR}).
When $f_1$ and $f_2$ interact with a vector boson $B_\alpha$ and with
a fermion $F$ through eq.~(\ref{jsoeh}),
$\sigma_L$ and $\sigma_R$ receive the contributions
in eqs.~(\ref{uirmd})--(\ref{nfhit}).
All these results are completely general --- they do not depend
on the values of the kinematical variables $p_1^2 = m_1^2$,
$p_2^2 = m_2^2$,
and $q^2$.
The finite loop integrals $a, b_1, b_2,$ and so on
in the general expressions for $\sigma_L$ and $\sigma_R$
are defined through eqs.~(\ref{hfurt})--(\ref{nfgty}).

\end{document}